\documentclass[twocolumn]{revtex4}
\usepackage{graphicx}
\usepackage{subfigure} 
\usepackage[latin1]{inputenc}
\usepackage[brazil]{babel}
\usepackage{amsmath}
\usepackage{setspace}
\usepackage{amssymb}
\usepackage{color}

\begin{document}

\title{Uma Introdução ao Controle do Caos em Sistemas
Hamiltonianos Quase Integráveis}
\author{Vilarbo da Silva Junior}
\email{vilarbos@unisinos.br}
\author{Alexsandro M. Carvalho}
\email{alexsandromc@unisinos.br}
\affiliation{Centro de Ciências Exatas e Tecnológicas, Universidade do Vale do Rio dos Sinos, Caixa Postal 275, 93022-000 São Leopoldo RS, Brazil}

\begin{abstract}
Sistemas Hamiltonianos quase integráveis são de grande interesse em diversos campos de pesquisa da física e da matemática. Nestes sistemas, o espaço de fase apresenta trajetórias regulares e caóticas. Essas trajetórias dependem, em parte, da amplitude da perturbação que quebra a integrabilidade do sistema. O valor da perturbação crítica responsável por esta transição é um elemento chave no controle do caos.

No presente trabalho, exploramos um procedimento para o controle do caos em sistema hamiltoniano quase integrável via mapa canônico. Inicialmente, apresentamos as ferramentas básicas para este estudo: mapa hamiltoniano, linearização do mapa e critério de Chirikov. Posteriormente, investigamos o comportamento de uma interação do tipo onda-partícula frente a perturbação. Por fim, confrontamos os resultados analíticos com uma abordagem numérica (iteração do mapa), mostrando um bom acordo.

\end{abstract}
\maketitle

\section{Introdução}

Tópicos relacionados à sistemas Hamiltonianos quase integráveis são importantes em diversos campos de pesquisa, tais como: sistemas dinâmicos~\cite{Kathleen}, física estatística~\cite{Dorfman}, física da matéria condensada~\cite{Aoki}, mecânica quântica semiclássica~\cite{Micklitz} e  física de plasma~\cite{Shuklaa,rizzato1, rizzato2}. Por exemplo, na mecânica clássica, embora se possa geralmente formular equações de movimento para um sistema arbitrariamente complexo que descrevam completamente a sua dinâmica, essas equações resultantes não são, em geral, exatamente solúveis. Neste caso, a técnica de perturbação~\cite{Nivaldo} é um recurso analítico relevante.

Ao pertubar um sistema Hamiltoniano, verificamos que uma perturbação não nula e pequena pode destruir parcialmente os toros ressonantes, apenas uma parte dos toros sobrevivem distorcidos. Neste caso, dizemos que o sistema é quase integrável, na medida em que o espaço de fase é ainda povoado por toros sobre os quais temos trajetórias com as mesmas características dos sistemas integráveis. É de interesse saber quais destes toros são destruídos e quais sobrevivem. De acordo com o teorema KAM~\cite{kolmogorov,arnold2,moser1}, os toros que sobrevivem são aqueles que têm um quociente de frequência suficientemente irracional.

Entretanto, gostaríamos de ter um critério preciso para essa transição. Um critério útil e intuitivo foi formulado por Chirikov~\cite{chirikov}. Na sua formulação mais simples, o critério diz que o caos global se inicia quando a intensidade da perturbação é suficientemente grande para que a diferença na ação de ressonâncias vizinhas seja comparável à largura na ação das ilhas de estabilidade. Nesta situação, a partícula passa a ``viajar'' por todo o espaço de fase e órbitas caóticas se formam.

Neste trabalho examinamos um procedimento para investigar o caos em sistemas Hamiltonianos quase integráveis.  Em particular, exploramos uma versão simplificada para o modelo de uma partícula relativística movendo-se sob a influência de um campo magnético uniforme e onda estacionária eletrostática~\cite{rizzato1,rizzato2}.

Este trabalho está organizado como segue: Na seção~\ref{aspectos} revisamos alguns conceitos básicos na investigação do caos em sistemas hamiltonianos. Primeiro, indicamos como converter as equações de Hamilton a um mapa e, posteriormente, como linearizá-lo. Na sequência, apresentamos um método que estabelece um parâmetro físico para o início do movimento caótico em sistemas hamiltonianos determinísticos, critério de Chirikov. Baseado no exemplo da interação onda-partícula, seção~\ref{exemplo}, indicamos como as ferramentas apresentadas na seção anterior são úteis no controle do caos. Para tanto, determinamos expressões analíticas para localização das ressonâncias primárias bem como para o valor crítico da perturbação. Adicionalmente, confirmamos estas soluções pela implementação numérica do mapa. Por fim, na seção~\ref{conclusao} realizamos nossas considerações finais.

\section{Aspectos Gerais}
\label{aspectos}
Caos aparece como consequência natural da não integrabilidade das equações de Hamilton para sistemas com mais de um grau de liberdade. Nesta seção apresentaremos alguns conceitos preliminares para o estudo do caos em sistemas hamiltonianos.

\subsection{Mapa Hamiltoniano}
\label{mapahamiltoniano}

Mapas são importantes ferramentas no estudo da dinâmica da evolução temporal de sistemas dinâmicos por serem de fácil implementação computacional. Para tanto, considere um Hamiltoniano $H(q,p)$ em que $q$ e $p$ representam as coordenadas de posição e \textit{momentum} generalizados, respectivamente. Assim, $H$ satisfaz as equações de Hamilton
\begin{eqnarray}
\dot{q}&=&\dfrac{\partial H}{\partial p} \\
\dot{p}&=&-\dfrac{\partial H}{\partial q}.
\end{eqnarray}
Agora, escrevemos
\begin{equation}
\dot{q_n} = \frac{q_{n+1}-q_n}{\Delta t}
\end{equation}
em que $q_n=q(t)$ e $q_{n+1}=q(t+\Delta t)$.
Sendo assim, as equações de movimento tornam-se
\begin{eqnarray}
q_{n+1}&=&q_{n}+\Delta t\left(\dfrac{\partial H}{\partial p}\right)_{(q_n,p_n)} \\
p_{n+1}&=&p_{n}-\Delta t\left(\dfrac{\partial H}{\partial q}\right)_{(q_n,p_n)}.
\end{eqnarray}

As equações anteriores correspondem ao mapeamento que fornecem $q$ e $p$ no instante $n+1$, conhecidos seus valores no instante $n$.
Vale salientar que dependendo da forma do Hamiltoniano, o sistema anterior não preserva a área $\partial (q_{n+1},p_{n+1})/\partial (q_n,p_n)\neq 1 $ (determinante do jacobiano diferente da unidade). Assim, para preservar a área, por vezes, é interesse realizar $(q_n,p_n) \rightarrow (q_{n+1},p_n)$ na segunda equação do mapa. Para mais detalhes veja o trabalho de Lichtenberg~\cite{Lichtenberg}.

\subsection{Linearização do Mapa}
\label{lineargeral}

Por vezes é útil linearizar um sistema dinâmico a tempo discreto (mapa) em torno de seus pontos fixos~\cite{artur}.

Consideremos o caso de um mapa bidimensional $\textbf{T}:\mathbb{R}^2\rightarrow \mathbb{R}^{2}$ definido por
\begin{eqnarray}
\label{eq37}
x_{n+1}&=&f(x_{x},y_{n})\\ y_{n+1}&=&g(x_{x},y_{n}),
\label{eq38}
\end{eqnarray}
ou, em notação compacta
\begin{equation}
\textbf{x}_{n+1}=\textbf{T}(\textbf{x}_{n}),
\label{eq38}
\end{equation}
onde $\textbf{x}_{n}=(x_{n},y_{n})$ e $\textbf{T}(\textbf{x}_{n})=(f(x_{x},y_{n}),g(x_{x},y_{n}))$.

Um ponto $\textbf{x}_{o}\in \mathbb{R}^{2}$ é dito de equilíbrio, ou \textit{fixo} para a dinâmica gerada por $\textbf{T}$ se, $\forall\, k\in
\mathbb{N}$ vale a condição
\begin{equation}
\textbf{T}^{k}(\textbf{x}_{0})=\textbf{x}_{0}.
\label{eq39}
\end{equation}

Assim, o mapeamento linearizado em torno de um ponto fixo $\textbf{x}_{0}$ tem a forma
\begin{equation}
\textbf{z}_{n+1}=\left(
                   \begin{array}{cc}
                     f_{x_{n}}(\textbf{x}_{0}) & f_{y_{n}}(\textbf{x}_{0}) \\
                     g_{x_{n}}(\textbf{x}_{0}) & g_{y_{n}}(\textbf{x}_{0}) \\
                   \end{array}
                 \right)\textbf{z}_{n},
\label{eq40}
\end{equation}
onde os subscritos em cada entrada da matriz Jacobiana indicam as respectivas derivadas parciais. De forma compacta, escrevemos a eq.~(\ref{eq40}) como
$\textbf{z}_{n+1}=A\textbf{z}_{n}$. Esta forma linearizada do mapa $\textbf{T}$ permite-nos empregar (localmente) as técnicas de
estabilidade linear. Em outras palavras, analisamos a estabilidade dos pontos fixos da equação anterior por meio de um estudo dos autovalores da matriz
Jacobiana. Posteriormente, recordamos que a equação característica associada a matriz $A$  pode ser expressa na forma $\lambda^{2}-\lambda\,
tr(A)+\det{(A)}=0$, onde $\lambda$ é um autovalor de $A$, $tr(A)$ seu traço e $\det{(A)}$ seu determinante.

\subsection{Critério de Chirikov}
\label{chirikov}
O critério Chirikov, foi introduzida em 1959 por Boris Chirikov \cite{chirikov2} e aplicado com sucesso para explicar a
fronteira de confinamento de plasma em ``armadilhas'' observadas em experimentos no Instituto Kurchatov.

De acordo com este critério, uma trajetória determinística começará a mover-se entre duas ressonâncias não lineares de uma maneira caótica e
imprevisível, logo que estas ressonâncias se sobrepõem. Isto ocorre quando a perturbação ou parâmetro de caos torna-se maior do que um determinado valor
($\epsilon_{c}$). Desde a sua introdução, o critério Chirikov tornou-se uma importante ferramenta analítica para a determinação da fronteira do caos em
sistemas Hamiltonianos. No que segue, iremos esboçar a teoria que arquiteta tal critério.

Vamos supor que um Hamiltoniano genérico (fracamente) perturbado possa ser escrito na forma
\begin{equation}
H(J,\theta,t)=H_{o}(J)+\epsilon H_{1}(J,\theta,t),
\label{eq18}
\end{equation}
onde $J$ e $\theta$ são variáveis de ação e ângulo para um sistema não-perturbativo de um grau de liberdade, mas periódico no tempo e que $\epsilon$ é
relativamente pequeno.

Por ser $\theta$ uma variável de ângulo, segue que $H_{1}(J,\theta,t)$ é uma função periódica de $\theta$ com um certo período $T_{\theta}$. Vamos supor
também que seja uma função periódica do tempo com período $T_{\lambda}$ (e assim com frequência $\omega_{\lambda}=2\pi/T_{\lambda}$). Deste modo, $H_{1}(J,\theta,t)$ pode ser expandido em uma série dupla de Fourier, ou seja,
\begin{equation}
H_{1}(J,\theta,t)=\sum_{m,n\in \mathbb{Z}}(H_{1}(J))_{m,n}e^{i\Omega},
\label{eq19}
\end{equation}
onde $\Omega=m\theta+n\omega_{\lambda}t$ e
\begin{equation}
(H_{1}(J))_{m,n}=\frac{1}{T_{\theta}T_{\lambda}}\int_{0}^{T_{\theta}}\int_{0}^{T_{\lambda}}H_{1}(J,\theta,t)e^{-i\Omega}d\theta d t.
\label{eq20}
\end{equation}

A condição de ressonância é dada por
\begin{equation}
\frac{d}{dt}(\overline{m}\,\theta+\overline{n}\,\omega_{\lambda}t)=0 \leftrightarrow \overline{m}\,\omega_{o}+\overline{n}\,\omega_{\lambda}=0,
\label{eq21}
\end{equation}
para algum par $(\overline{m},\overline{n})\in \mathbb{Z}^{2}$ e
\begin{equation}
\omega_{o}(\overline{m},\overline{n})=\frac{\partial H_{o}(J_{\overline{m},\overline{n}})}{\partial J}.
\label{eq22}
\end{equation}
Evidenciamos que primeiro devemos derivar $H_{o}(J)$ com respeito à ação para depois avaliar seu valor no particular ponto
$J_{\overline{m},\overline{n}}$ (valor da ação para o qual vale a condição de ressonância, eq.~(\ref{eq21})).

Afim de realizar uma transformação canônica dependente do tempo, tomamos a função geratriz
\begin{equation}
F_{2}(\theta,I,t)=(I+J_{\overline{m},\overline{n}})(\theta-\omega_{o}(\overline{m},\overline{n}) t).
\label{eq23}
\end{equation}
Deste modo, pela teoria das transformações canônicas~\cite{walter} segue que
\begin{eqnarray}
J&=&\frac{\partial F_{2}}{\partial \theta}=I+J_{\overline{m},\overline{n}}\\
\label{eq24}
\vartheta &=&\frac{\partial F_{2}}{\partial I}=\theta-\omega_{o}(\overline{m},\overline{n}) t.
\label{eq25}
\end{eqnarray}

Assim, as novas variáveis de ação e ângulo ficam
\begin{equation}
I=J-J_{\overline{m},\overline{n}};\quad \vartheta =\theta-\omega_{o}(\overline{m},\overline{n}) t.
\label{eq26}
\end{equation}

Portanto o novo Hamiltoniano $K(\vartheta,I,t)$ deve ser escrito, utilizando as equações~(\ref{eq18}), (\ref{eq19}), (\ref{eq23}) e (\ref{eq26}), como
\begin{eqnarray}
K(\vartheta,I,t)&=&H+\frac{\partial F_{2}}{\partial t}=\nonumber\\
&=&H_{o}(J_{\overline{m},\overline{n}})-\omega_{o}(\overline{m},\overline{n})J_{\overline{m},\overline{n}}+\frac{I^{2}}{2}\frac{\partial^{2}H_{o}(J_{\overline{m},\overline{n}})}{\partial
I^{2}}\nonumber\\ &+&\epsilon 2\,(H_{1}(J_{\overline{m},\overline{n}}))_{\overline{m},\overline{n}}\cos{(\overline{m}\,\vartheta)},
\label{eq27}
\end{eqnarray}
onde para o desenvolvimento da expressão~(\ref{eq27}) realizamos:
\begin{enumerate}
\item  expansão de Taylor ($I=J-J_{\overline{m},\overline{n}}\ll 1$)
\item  expansão de Fourier (termos ressonantes)
\item  $(H_{1})_{-\overline{m},-\overline{n}}=(H_{1})_{\overline{m},\overline{n}}$ (coeficientes reais)
\item  $\cos{(x)}=(e^{ix}+e^{-ix})/2$
\end{enumerate}

Posteriormente, definimos o Hamiltoniano Ressonante
\begin{equation}
h^{r}_{\overline{m},\overline{n}}:=K(\vartheta,I,t)-H_{o}(J_{\overline{m},\overline{n}})+\omega_{o}(\overline{m},\overline{n})J_{\overline{m},\overline{n}},
\label{eq28}
\end{equation}
a massa efetiva
\begin{equation}
M^{-1}:=\frac{\partial^{2}H_{o}(J_{\overline{m},\overline{n}})}{\partial I^{2}}
\label{eq29}
\end{equation}
e a intensidade de ressonância
\begin{equation}
\Lambda_{\overline{m},\overline{n}}:=2\epsilon(H_{1}(J_{\overline{m},\overline{n}}))_{\overline{m},\overline{n}}.
\label{eq30}
\end{equation}

Em síntese, todo este procedimento nos leva à forma
\begin{equation}
h^{r}_{\overline{m},\overline{n}}=\frac{1}{2M}I^{2}+\Lambda_{\overline{m},\overline{n}}\cos{(\overline{m}\,\vartheta)}.
\label{eq31}
\end{equation}

O Hamiltoniano efetivo~(\ref{eq31}) controla a dinâmica nas imediações de uma ressonância caracterizada pela relação~(\ref{eq21}). A ação $I$
que aparece na eq.~(\ref{eq31}) representa a flutuação da ação em torno de $J_{\overline{m},\overline{n}}$.

A separatriz corresponde a tomar $h^{r}_{\overline{m},\overline{n}}=\Lambda_{\overline{m},\overline{n}}$ o que conduz-nos a
$J_{max}=2\sqrt{M\,\Lambda_{\overline{m},\overline{n}}}$ e assim limita regiões do espaço de fases com largura máxima
\begin{equation}
\Delta J=\sqrt{\frac{32\epsilon
(H_{1}(J_{\overline{m},\overline{n}}))_{\overline{m},\overline{n}}}{\left|\frac{\partial^{2}H_{o}(J_{\overline{m},\overline{n}})}{\partial
I^{2}}\right|}}.
\label{eq32}
\end{equation}

Se $J_{\overline{m}+1,\overline{n}}$ é a próxima ressonância estima-se, para $\overline{m}$ suficientemente grande, que
\begin{equation}
\delta:=|J_{\overline{m}+1,\overline{n}}-J_{\overline{m},\overline{n}}|\approx
\frac{\omega_{\lambda}}{\overline{m}^{2}}\left|\frac{\partial^{2}H_{o}(J_{\overline{m},\overline{n}})}{\partial I^{2}}\right|^{-1},
\label{eq33}
\end{equation}
onde a condição $\delta=\Delta J$ fornece o valor crítico de $\epsilon_{c}$ para a superposição de regiões $m$-ressonantes. Mais
especificamente,
\begin{equation}
\epsilon_{c}=\frac{\omega_{\lambda}^{2}}{32\overline{m}^{4}(H_{1}(J_{\overline{m},\overline{n}}))_{\overline{m},\overline{n}}}\left|\frac{\partial^{2}H_{o}(J_{\overline{m},\overline{n}})}{\partial
I^{2}}\right|^{-1}.
\label{eq34}
\end{equation}

Esse é um critério aproximado para que o campo crítico destrua as superfícies KAM entre a $\overline{m}$-ésima e a $(\overline{m}+1)$-ésima ilha,
possibilitando que a partícula excursione livremente de uma ressonância clássica a outra. Para $\overline{m}$ maior a eq.~(\ref{eq34}) mostra que $\epsilon_{c}$ é menor e, portanto, há mais caos.

\section{Exemplo: Interação Onda-Partícula}
\label{exemplo}

A interação onda-partícula aparece no estudo de problemas, tais como: aceleradores de partículas e lasers de elétrons livres (para mais exemplos veja o trabalho de Souza et al.~\cite{rizzato1}). Este tipo de interação resulta em um processo não linear que pode apresentar trajetórias regulares e caóticas no espaço de fase. A presença de uma ou outra trajetória depende principalmente da amplitude de perturbação aplicada ao sistema.

\subsection{Modelo}
\label{secaomodelo}

Consideramos uma partícula relativística de carga $e$, massa $m$ e momento canônico $\textbf{p}$ que se move sobre a ação combinada de um campo magnético uniforme $\textbf{B}=B_{o}\textbf{k}$ e de uma ``onda'' eletrostática da forma
\begin{equation}
U(x,t)=\varepsilon x^{2} \sum_{k=-\infty}^{\infty}\delta (t-kT_{\lambda}),
\label{eq1}
\end{equation}
onde $T_{\lambda}$ é o período de modulação e $\varepsilon$ é a intensidade dos impulsos (perturbação) ao longo do eixo $x$. Desta maneira, toda vez que $t=kT_{\lambda}$ com $k\in \mathbb{Z}$, a perturbação é ativada por um instante infinitesimalmente pequeno.

Deste modo, a dinâmica transversal deste sistema é descrita pelo Hamiltoniano~\cite{rizzato1,rizzato2}
\begin{equation}
H=\sqrt{m^{2}c^{4}+c^{2}p_{x}^{2}+c^{2}(p_{y}+eB_{o}x)^{2}}+\varepsilon U(x,t),
\label{eq2}
\end{equation}
onde $c$ é a velocidade da luz. Note que o Hamiltoniano não depende da variáveis $y$ (variável cíclica) e, consequentemente, $p_{y}$ é uma constante de movimento. Assuminos sem perda de generalidade que $p_{y}=0$. Destacamos que embora $p_{y}$ seja nula, $dy/dt$ é diferente de zero.

Indo de encontro a reescrever o Hamiltoniano~(\ref{eq2}) em termos de quantidades adimensionais, definimos as seguintes mudanças de variáveis
$H/mc^{2}\rightarrow \mathcal{H}$, $p_{x}/mc\rightarrow p$, $eB_{o}x/mc\rightarrow q$ e $e^{2}B_{o}^{2}\varepsilon/m\rightarrow \epsilon$. Com efeito,
\begin{equation}
\mathcal{H}=\sqrt{1+p^{2}+q^{2}}+\epsilon\, q^{2} \sum_{k=-\infty}^{\infty}\delta (t-kT_{\lambda}).
\label{eq3}
\end{equation}
É conveniente escrever o Hamiltoniano anterior em termos das variáveis de ângulo e ação que diagonalize a parte não perturbada. Para tanto,
consideramos a função geratriz~\cite{Nivaldo} $S(I,q)$ que gera a seguinte mudança de coordenadas
\begin{equation}
p=\sqrt{2I}\cos{(\theta)}; \qquad\qquad q=\sqrt{2I}\sin{(\theta)},
\label{eq4}
\end{equation}
onde $I$ e $\theta$ são as variáveis de ação e ângulo, respectivamente.

Substituindo a  eq.~(\ref{eq4}) na eq.~(\ref{eq3}) chegamos em
\begin{equation}
\mathcal{H}(I,\theta,t)=\sqrt{1+2I}+\epsilon\, 2 I \sin^{2}{(\theta)}\sum_{k=-\infty}^{\infty}\delta (t-kT_{\lambda}).
\label{eq5}
\end{equation}

Observe que o Hamiltoniano acima está na forma $\mathcal{H}=H_{o}(I)+\epsilon H_{1}(I,\theta,t)$, onde
\begin{equation}
H_{o}(I)=\sqrt{1+2I},
\label{eq6}
\end{equation}
e
\begin{equation}
H_{1}(I,\theta,t)=2 I \sin^{2}{(\theta)}\sum_{k=-\infty}^{\infty}\delta (t-kT_{\lambda}).
\label{eq7}
\end{equation}

Perceba que $H_{o}$ só depende da variável de ação $I$, e que $H_{1}$ é uma função periódica do tempo, com período $T_{\lambda}$. Por conseguinte,
podemos associar uma frequência temporal $\omega_{\lambda}=2\pi/T_{\lambda}$ ao termo perturbado. Além disso, a ``onda'' depende funcionalmente de
$\sin^{2}{(\theta)}$, deste modo sua periodicidade básica é $\pi$.

\subsection{Ressonâncias}
\label{LocalRessonan}

Afim de localizar as ressonâncias primárias do sistema, devemos escrever o Hamiltoniano (\ref{eq7}) como uma dupla expansão de Fourier tanto no tempo
quanto na variável angular. Assim, levando em conta que $\sin^{2}{(\theta)}=(2-e^{2i\theta}-e^{-2i\theta})/4$ e que
$\sum_{k=-\infty}^{\infty}\delta(t-kT_{\lambda})=\sum_{n=-\infty}^{\infty}e^{in\omega_{\lambda}t}/T_{\lambda}$ segue que o Hamiltoniano (\ref{eq5}) assume
a forma
\begin{equation}
\mathcal{H}(I,\theta,t)=\sqrt{1+2I}+\epsilon\frac{I}{T_{\lambda}}\Theta(\theta,\omega_{\lambda}),
\label{eq13}
\end{equation}
onde
\begin{equation}
\Theta(\theta,\omega_{\lambda})=\sum_{n=-\infty}^{\infty}\left[e^{in\omega_{\lambda}t}-\frac{1}{2}\left(e^{i\triangle_{-}}+e^{i\triangle_{+}}\right)\right],
\label{eq14}
\end{equation}
onde $\triangle_{\pm}:=\pm 2\theta+n\omega_{\lambda}t$.

As expressões (\ref{eq13})-(\ref{eq14}) nos permite detectar a presença de ressonâncias. As ressonâncias estão localizadas, como visto na seção
(\ref{chirikov}), em particulares valores $m, n$, os quais denotaremos por $\overline{m}$ e $\overline{n}$ tais que
$d(\overline{m}\theta+\overline{n}\omega_{\lambda}t)/dt=0$. Considerando a condição de ressonância, concluímos que as mesmas
encontram-se em
\begin{eqnarray}
\label{eq15}
    0\omega_{o}(I)+\overline{n}\omega_{\lambda}&=&0, \\
\label{eq16}
    2\omega_{o}(I)+\overline{n}\omega_{\lambda}&=&0,\\
    -2\omega_{o}(I)+\overline{n}\omega_{\lambda}&=&0.
\label{eq17}
\end{eqnarray}
onde a frequência $\omega_{o}(I)$ deve ser calculada como na eq.~(\ref{eq22}). Assim, resulta que
\begin{equation}
\omega_{o}(I)=\frac{\partial H_{o}}{\partial I}=\frac{1}{\sqrt{1+2I}}.
\label{eq35}
\end{equation}

Observando as eq.~(\ref{eq15})-(\ref{eq17}) vemos que existe, em princípio, três valores possíveis para $\overline{m}$ ($\overline{m}=0$,
$\overline{m}=2$ e $\overline{m}=-2$). Entretanto, pela forma funcional de $\omega_{o}(I)$, os dois últimos valores de $\overline{m}$ são
equivalentes e assim as ressonâncias serão encontradas variando $\overline{n}$. Da eq.~(\ref{eq15}) segue que $\overline{n}=0$. Decorre, para as demais ressonâncias
\begin{equation}
I_{2,\overline{n}}=\frac{1}{2}\left(\frac{4}{(\overline{n}\omega_{\lambda})^{2}}-1\right)
\label{eq36}
\end{equation}

Por exemplo, se $T_{\lambda}=2.5\pi$  ($\omega_{\lambda}=4/5$) para $\overline{n}=1$ obtemos $I_{2,1}=2.625$. Analogamente, para $\overline{n}=2$ encontramos $I_{2,2}=0.281$. Todas estes valores
de $I$ ocorrem em $\theta=\pi/2$ e $\theta=3\pi/2$, pontos fixos. Assim, as ressonâncias estão
localizadas em $(\pi/2,2.625)$, $(3\pi/2,2.625)$, $(\pi/2,0.281)$ e $(3\pi/2,0.281)$ (notação $(\theta, I)$). Além disso, determinamos  $\epsilon_{c}=0.1364$, valor crítico para o qual órbitas iniciando nas vizinhanças de $I_{2,2}$ comecem a
migrar para a ressonância $I_{2,1}$.

\subsection{Mapa}
\label{sectmapeamentp}

\begin{figure}[h]
\centerline{
\includegraphics[width=0.5\textwidth]{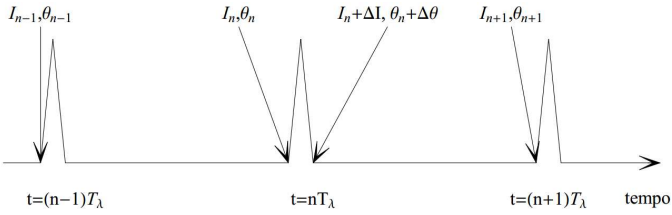}}
\caption{Esquema para obtenção do mapeamento associado ao Hamiltoniano (\ref{eq5}).}
\label{fig1}
\end{figure}

A dinâmica gerada pelo Hamiltoniano~(\ref{eq5}) pode ser parcialmente integrada. Para tal, fazemos uso do esquema apresentado na Fig.~\ref{fig1}. Sendo assim, $I_{n}$ e $\theta_{n}$ são os valores de $I$ e $\theta$ na ``entrada'' (logo a esquerda) da função delta correspondendo a
$t=nT_{\lambda}$. Assim, podemos construir analiticamente a conexão genérica entre $(\theta_{n},I_{n})$ e $(\theta_{n+1},I_{n+1})$.
Com efeito, se denotarmos por $\Delta \theta$ e $\Delta I$ as variações de $\theta$ e $I$ após a travessia de uma função delta centrada em $t=nT_{\lambda}$ teremos, baseado na subseção~\ref{mapahamiltoniano},
\begin{equation}
\Delta \theta = \int_{nT_{\lambda-0}}^{nT_{\lambda+0}}\dot{\theta}\,dt=\frac{T_{\lambda}}{\sqrt{1+2I_{n+1}}}+2\epsilon \sin^{2}{(\theta_{n})},
\label{eq8}
\end{equation}
e
\begin{equation}
\Delta I = \int_{nT_{\lambda-0}}^{nT_{\lambda+0}}\dot{I}\,dt=-2\epsilon I_{n+1}\sin{(2\theta_{n})}.
\label{eq9}
\end{equation}

Portanto,
\begin{eqnarray}
\label{eq10}
\theta_{n+1}&=&\theta_{n}+\frac{T_{\lambda}}{\sqrt{1+2I_{n+1}}}+2\epsilon \sin^{2}{(\theta_{n})}, \\
    I_{n+1}&=&\frac{I_{n}}{1+2\epsilon\sin{(2\theta_{n})}}.
\label{eq11}
\end{eqnarray}
Note que o mapa pode ser interpretado como uma transformação canônica \cite{Nivaldo} entre as
variáveis ``velhas'' $(\theta_{n},I_{n})$ e as ``novas'' $(\theta_{n+1},I_{n+1})$, gerada pela seguinte função geratriz
\begin{eqnarray}
G(\theta_{n},I_{n+1})&=&\theta_{n}I_{n+1}+T_{\lambda}\sqrt{1+2I_{n+1}}\nonumber\\ &+&2\epsilon I_{n+1}\sin^{2}{(\theta_{n})},
\label{eq12}
\end{eqnarray}
onde $\theta_{n+1}=\partial G/\partial I_{n+1}$ e $I_{n}=\partial G/\partial \theta_{n}$.

Cabe observar que o mapa poderia ter sido escrito com o mesmo grau de complexidade nas variáveis $(q,p)$. No entanto, o uso das
variáveis de ângulo e ação $(\theta,I)$ são mais conveniente posto que a ação é conservada na ausência de perturbação.

\subsection{Linearização do Mapa}
\label{linearsystem}

Aplicaremos as técnicas apresentadas na subseção~\ref{lineargeral} ao caso particular do mapa definido anteriormente, afim de estudar a
estabilidade das ressoâncias. Desta maneira, segue que os elementos da matriz Jacobiana são
\begin{eqnarray}
\label{eq41}
\frac{\partial \theta_{n+1}}{\partial \theta_{n}}&=&1+2\epsilon \sin{(2\theta_{n})}-\frac{\partial I_{n+1}}{\partial
\theta_{n}}\frac{T_{\lambda}}{\sqrt{\beta_{n}^{3}}},\\
\label{eq42}
\frac{\partial \theta_{n+1}}{\partial I_{n}}&=&-\frac{\partial I_{n+1}}{\partial I_{n}}\frac{T_{\lambda}}{\sqrt{\beta_{n}^{3}}},\\
\label{eq43}
\frac{\partial I_{n+1}}{\partial \theta_{n}}&=&-\frac{4\,\epsilon I_{n}\cos{(2\theta_{n})}}{(1+2\,\epsilon \sin{(2\theta)})^{2}},\\
\label{eq44}
\frac{\partial I_{n+1}}{\partial I_{n}}&=&\frac{1}{1+2\,\epsilon \sin{(2\theta)}},
\end{eqnarray}
onde $\beta_{n}:=1+2I_{n+1}$.

Agora, vamos nos concentar na ressonância principal do mapa na
qual cada ciclo da onda corresponde a uma rotação orbital completa das partículas magnetizadas. Tal ressonância está localizada no ponto fixo angular
$\theta=\pi/2$, obtido com a condição $\theta_{n+1}-\theta_{n}=2\pi$, que define a posição da ressonância principal
\begin{equation}
I_{res}=\frac{T_{\lambda}^{2}-4(\pi-\epsilon)^{2}}{8(\pi-\epsilon)^{2}}.
\label{eqiresson}
\end{equation}
Ao substituírmos $\theta=\pi /2$ e $I_{n}=I_{res}$ nos elementos da matriz Jacobiana acima, somos conduzidos ao mapa linearizado
$\textbf{z}_{n+1}=A\textbf{z}_{n}$, onde
\begin{equation}
A=\left(
    \begin{array}{cc}
      1-\frac{4\epsilon T_{\lambda}I_{res}}{(1+2I_{res})^{3/2}} & -\frac{T_{\lambda}}{(1+2I_{res})^{3/2}} \\
      4\epsilon I_{res} & 1 \\
    \end{array}
  \right).
\label{eq45}
\end{equation}

Os autovalores de $A$ são obtidos como solução da equação característica
\begin{equation}
\lambda^{2}-2\lambda\Gamma+1=0,
\label{eq46}
\end{equation}
onde $\Gamma:=1-\frac{2\epsilon T_{\lambda}I_{res}}{(1+2I_{res})^{3/2}}$. A solução da eq.~(\ref{eq46}) nos leva a
\begin{equation}
\lambda_{1,2}=\Gamma\pm\sqrt{\Gamma^{2}-1}.
\label{eq47}
\end{equation}
Note que $\det(A)=1$, o que evidencia a propriedade de preservação da área no espaço de fase.

Nosso objetivo é determinar o valor de $\epsilon$ a partir do qual um dos autovalores se torna maior do que $1$. Neste instante o ponto
elíptico estável centrado, por exemplo, em $\pi/2$ se torna um ponto hiperbólico instável. Afim de realizar esta estimativa para o parâmetro de perturbação $\epsilon$, devemos impor a condição de que o argumento da raiz quadrada na eq.~(\ref{eq47}) seja maior do que zero. Este procedimento nos conduz a
\begin{equation}
\epsilon>\frac{(1+2I_{res})^{3/2}}{T_{\lambda}I_{res}},
\label{eq48}
\end{equation}
onde $I_{res}$ depende de $\epsilon$ via eq.~(\ref{eqiresson}).

Por exemplo, para $T_{\lambda}=2.5 \pi$ estima-se $\epsilon > 0.67$.

\subsection{Resultados Numéricos}
\label{numericalresults}

\begin{figure}[t!]
\centerline{
\includegraphics[width=8.0cm]{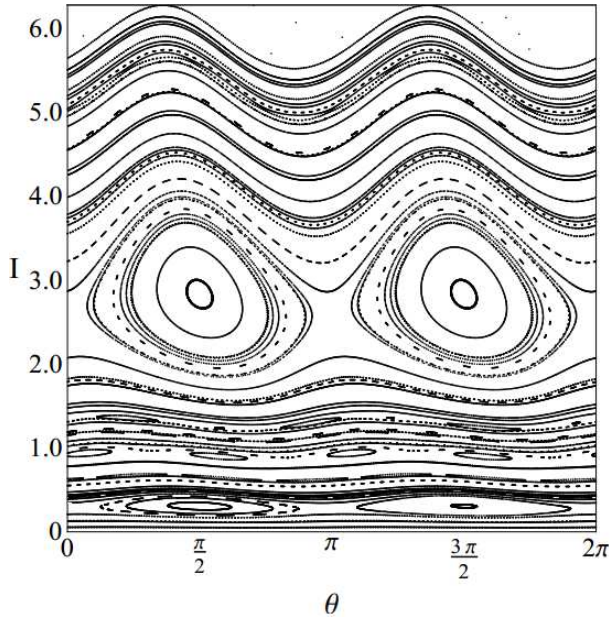}}
\caption{Espaço de fases associado ao mapa não linear, equações~(\ref{eq10}) e (\ref{eq11}). Implementado com $T_{\lambda}=2.5\pi$ e $\epsilon=0.05$.}
\label{fig2}
\end{figure}

\begin{figure}[t!]
\center
\subfigure[ref1][\hspace{0.3cm} $T_{\lambda}=2.5\pi$ e
$\epsilon=0.05$.]{\includegraphics[width=7cm]{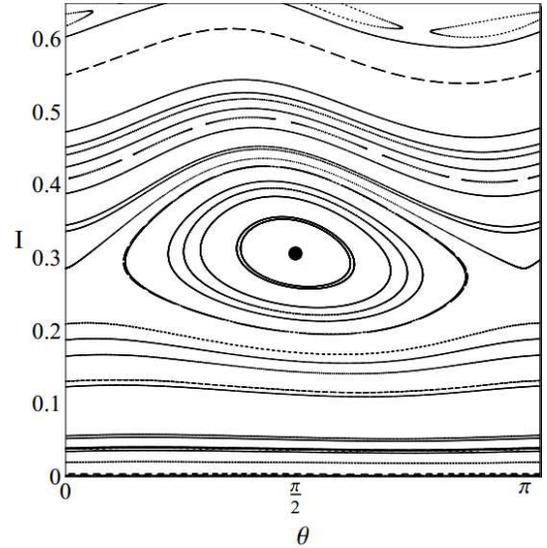}}
\qquad
\subfigure[ref2][\hspace{0.3cm} $T_{\lambda}=2.5\pi$ e
$\epsilon=0.15$.]{\includegraphics[width=7cm]{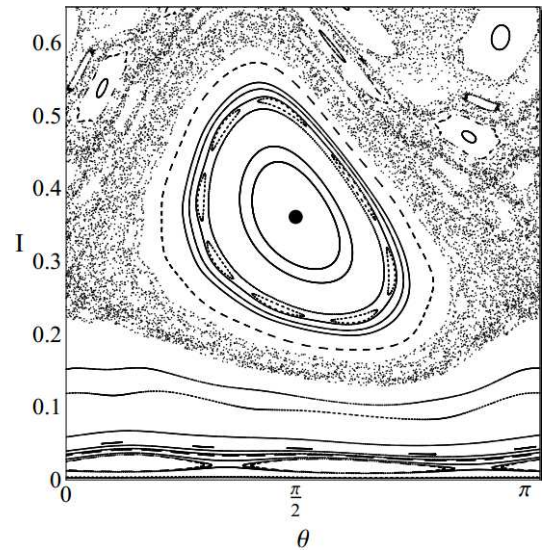}}
\caption{Janela $\{(\theta,I) \in [0, \pi]\times[0,0.6]\}$ associado ao mapa não linear, equações~(\ref{eq10}) e (\ref{eq11}).}
\label{fig3}
\end{figure}

\begin{figure}[t!]
\centerline{
\includegraphics[width=8.0cm]{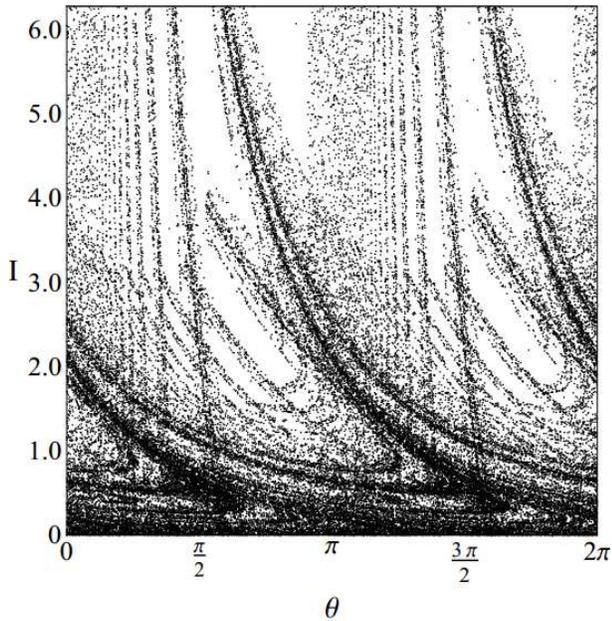}}
\caption{Plano de fases associado ao mapa não linear , eq.~(\ref{eq10}) e (\ref{eq11}). Implementado com $T_{\lambda}=2.5\pi$ e $\epsilon=0.5$.}
\label{fig4}
\end{figure}

Nesta parte do trabalho implementamos o mapa definido pelas equações~(\ref{eq10}) e (\ref{eq11}).
Inicialmente, iteramos o mapa para $\epsilon = 0.05$ e $T_{\lambda}=2.5\pi>2\pi$ (maior do que a frequência de ciclotron), tal
solução numérica está apresentada na Fig.~\ref{fig2}. Desta figura, é possível verificar que as posições das ressonâncias primárias ($\epsilon$
relativamente pequeno) estão de acordo com nossas previsões analíticas feitas na subseção~\ref{LocalRessonan}. Em adição,
percebemos que para este valor de $\epsilon$ não há constatação de órbitas caóticas e assim a estrutura do espaço de fases é regular.

Entretanto, vimos na eq.~(\ref{eqiresson}) que o valor ressonante da variável de ação $I_{res}$ depende do parâmetro de perturbação $\epsilon$. Na Fig.~\ref{fig3}a destacamos a janela $\{(\theta,I) \in [0, \pi]\times[0,0.6]\}$ da Fig.~\ref{fig2}, onde marcamos com um ponto a correta localização da
ressonância principal $(\pi/2,0.307)$. Além disso, é interessante ressaltar que tal ressonância se apresenta como no caso de um pêndulo, sendo um ponto
fixo elíptico (portanto estável).

Na subseção~\ref{LocalRessonan} estimamos o valor crítico de $\epsilon$, a partir do qual órbitas de ressonância correspondeno a $\Delta
\theta=2\pi$ começam a migrar para as de ressonância $\Delta \theta = \pi$, como sendo $\epsilon_{c}=0.1364$. Motivados por este valor,
analisamos na Fig.~\ref{fig3}b a mesma janela $\{(\theta,I) \in [0, \pi]\times[0,0.6]\}$ analisada na Fig.~\ref{fig3}a para
$T_{\lambda}=2.5\pi$ e $\epsilon=0.15>\epsilon_{c}$. Destacamos com um ponto a correta localização da ressonância principal para este valor de $\epsilon$,
ou seja, em $(\pi/2,0.3615)$. Além disso, conforme se aumenta a intensidade da pertubação vai ocorrendo a destruição dos toros com frequências de razão
racionais e uma estrutura auto similar vai surgindo. Fica evidente a presença de zonas caóticas. Estas são não
conectadas, o que é devido a existência de superfícies KAM entre elas. Podemos ver que dentro das zonas caóticas existem ilhas sem caos.

Na subseção~\ref{linearsystem}, foi previsto que é necessário $\epsilon>0.67$ afim de que um dos autovalores da matriz~(\ref{eq45}) do mapa
linearizado se torne maior do que $1$ (implicando na instabilidade de alguma ressonância). Entretanto, após algumas experiências numéricas, constatamos que ao tratar diretamente o mapa não linear o maior valor admissível para $\epsilon$ foi $0.5$ e a implementação para este valor de
$\epsilon$ está apresentada na Fig.~\ref{fig4}. O problema computacional encontrado para $\epsilon>0.5$ foi que o mapeamento passou a retornar valores
não reais. Notamos que, neste caso, todo espaço de fases apresenta-se caótico e assim as ilhas não caóticas desapareceram.
\vspace{0.2cm}

\section{Conclusões}
\label{conclusao}

Em conclusão, analisamos e desenvolvemos um modelo que descreve a dinâmica de interação de uma partícula relativística com um campo magnético uniforme e
uma onda eletrotática dada como uma série de pulsos. Partindo do Hamiltoniano do sistema, apresentamos um procedimento para obtenção do mapa que descreve
sua evolução temporal. Realizamos algumas previsões analíticas, tais como, obtenção de um valor crítico $\epsilon$ a partir do qual dá-se início o
movimento caótico $(\epsilon_{c}=0.1364)$ e a localização das ressonância primárias. Ao linearizar o mapa, estimamos o valor de $\epsilon$ que torna
uma ressonância principal originalmente estável em instável ($\epsilon=0.67$).

Com o mapa obtido construimos o espaço de fases do sistema e analisamos seu comportamento como sendo regular ou caótico. Basedo nos espaços de fases vemos
que a natureza qualitativa do movimento depende da intensidade de amplitude da onda eletrostática ($\epsilon$). Verificamos, numericamente, a posição das
ressonâncias  e estas estão de acordo com as previsões analíticas. Na Fig.~\ref{fig2} apresentamos o espaço de fases para $\epsilon=0.05$, onde não há
percepção de comportamento caótico mas sim de uma estrutura relativamente regular. Na Fig.~\ref{fig3}a destacamos a janela $\{(\theta,I) \in
[0, \pi]\times[0,0.6]\}$ da Fig.~\ref{fig2}, onde fica claro que a ressonância marcada com um ponto se trata de um ponto fixo elíptico (estável).

Na subseção~\ref{LocalRessonan} estimamos que $\epsilon > \epsilon_{c}=0.1364$, a fim de que órbitas vivendo nas vizinhanças correspondendo a $\Delta
\theta=2\pi$ comecem a migrar para a ressonância correspondente a $\Delta \theta = \pi$.  Deste modo, analisamos na Fig.~\ref{fig3}b a mesma janela
$\{(\theta,I) \in [0, \pi]\times[0,0.6]\}$ destacada na Fig.~\ref{fig3}a para $\epsilon=0.15>\epsilon_{c}$. Assim, verificamos que conforme se
aumenta a intensidade da pertubação vai ocorrendo a destruição dos toros com frequências de razão racionais e uma estrutura auto similar vai
surgindo (figura~\ref{fig3}b ) além das zonas caóticas.

Finalmente, concluímos que o algoritmo empregado para gerar a iteração numérica do mapa é robusto e pode ser empregado para iterações de outros mapas
bidimensionais deste tipo. Entretanto, a implementação deste mapa específico apresentou limitações para $\epsilon>0.5$ o que nos impossibilitou verificar a estimativa de que $\epsilon>0.67$ afim de que um dos autovalores da matriz~(\ref{eq45}) do mapa linearizado se torne maior do que $1$ (implicando na
instábilidade da ressonância entrada em $\pi/2$). Interpretamos esta discrepância como sendo devida a não linearidade do mapa original (em contrapartida
ao mapa linearizado).

\section*{Agradecimentos}
Agradecemos ao Professor Felipe Barbedo Rizzato pelas discussões e sugestões e ao Professor Rogério Steffenon pelo frequente incentivo.

\bibliographystyle{phcpc} 

\end{document}